# DISTRIBUTED ACCELERATED PROJECTION-BASED CONSENSUS DECOMPOSITION

**Wiktor Maj**
Centre of Informatics – Tricity Academic Supercomputer Network (CI TASK)
Gdansk University of Technology
wiktor.maj@pg.edu.pl

## Abstract
With the development of machine learning and Big Data, the concepts of linear and non-linear optimization techniques are becoming increasingly valuable for many quantitative disciplines. Problems of that nature are typically solved using distinctive optimization algorithms, iterative methods, or heuristics. A new variant of the Accelerated Projection-Based Consensus (APC) iterative method is proposed, which is faster than its classical version while handling large sparse matrices in distributed settings. The algorithm is proposed, and its description and implementation in a high-level programming language are presented. Convergence tests measuring acceleration factors based on real-world datasets are done, and their results are promising. The results of this research can be used as an alternative to solving numerical optimization problems.

# 1. Introduction

Examining an overdetermined system of linear equations $Ax = b$, typically for a sizeable sparse coefficient matrix $A \in \mathbb{R}^{m \times n}$ with its corresponding constant terms vector $b \in \mathbb{R}^m$ where $m > n$, the goal is to find estimator $\hat{x}$ such that $\hat{x} = \arg\min_x ||b - Ax||$. Setting up Ordinary Least Squares (OLS) [1] coefficient vector of the least-squares hyperplane expressed as $\hat{x} = (A^T A)^{-1} A^T b$, where $A^T A$ is a Gram matrix [2], and $A^T b$ is known as the moment matrix of regressand by regressors [3], is one of the most recognized, well-performed technique to find unknown parameters. In a distributed environment, operating on compressed sparse matrices makes computation slow due to the inflexibility of arithmetic operations on them. On the other hand, decompressing complementary matrix $A$ and finding its inverses requires extensive use of computational resources, which can unschedule task running or even crash workers. Numerous large-scale optimization problems are unfeasible to be trained on a single host. For instance, language models quickly scale up out of memory boundaries due to the number of hyperparameters involved in training. Hundreds of DGX A100 multi-GPU servers have been used to train a 530-billion parameter model on text datasets [4]. Despite the additional complexity layer and potential performance bottlenecks, distributing the workflow across multiple computing nodes provides robustness and scalability to make the models run.

Two main approaches are adapted in distributed machine learning workflows: model parallelism and data parallelism. Usually, both of them are present in modern high-level programming frameworks. In the case of data parallelism, there is evidence supporting the efficacy of globally exemplary methods such as Distributed Gradient Descent (DGD) [5], Alternating Direction Method of Multipliers (ADMM) [6], and Accelerated Projection-Based Consensus (APC) [7] with a view of the growth in the popularity of Big Data. Moreover, the progression of the development of distributed algorithms has led to the rise of ample open-source, high-level distributed computing frameworks such as Apache Spark [8] and Dask [9], which organizations are successfully using for commercial applications. The preceding considerations have strongly influenced contributions to infrastructural changes from bare-metal to more portable, virtualized, or containerized paradigms.

The main contribution is the decomposition of Accelerated Projection-Based Consensus [7] (*See Acknowledgments*) to avoid computationally expensive operations, particularly matrix inversions. The idea is to put the largest number of small-sized tasks into processing instead of fewer, larger indivisible tasks to make a more performant use of parallelism (Section 2). The full necessary implementation (Section 3) in the Python [10] programming language is provided. Convergence behavior (Section 4) compared to classical APC is demonstrated, with speedup times to the referenced datasets. The manual example was conducted (Section 5), and the results of the experiments are interpreted (Section 6) with their explanation. The main focus was on linear and non-linear optimization problems. However, the underlying concepts are present in other research areas as well.

# 2. The Algorithm

For $I \in \mathbb{N}$ subsets of equations, let us find $A_i \subset A$, $i = 1, 2, ..., I$ such that $A_i \in \mathbb{R}^{m \times n}$ and $rank(A) = rank(A_i)$ with its corresponding $b_i \subset b$, $b_i \in \mathbb{R}^n$ constant terms vectors. For $m = n$, using the fact that the non-singular $A_i$ square matrix is invertible (pseudoinverses in modern programming frameworks use singular value decomposition [11], which slightly enlarges computational times), each system of equations $i$ consists of a single unique solution $x_i = A_i^{-1} b_i$ having $P_i = I_n - A_i^T (A_i A_i^T)^{-1} A_i$ projection matrix onto the nullspace of $A_i$ [7], where $I_n$ is a $(n \times n)$ identity matrix. In a variety of applications, symmetric matrices (meaning that $\langle A_i x, y \rangle = \langle x, A_i y \rangle \ \forall x, y \in \mathbb{R}^n$) appear intrinsically. In the case of $A_i$ matrix asymmetricity, it can be replaced by $A_i := 0.5(A_i + A_i^T)$ as a Toeplitz decomposition subproduct [12] assuming $A_i$ of no loss of generality [13]. Therefore, the orthonormal basis for the $null(A_i A_i^T) = null(A_i^T A_i)$ is what evinces that $P_i = I_n - A_i (A_i^T A_i)^{-1} A_i^T$.

Each matrix $A_i$ can be decomposed as $A_i = Q_i R_i$ using full QR factorization [14] where $Q_i$ is an orthogonal and $R_i$ is a square upper triangular matrix, respectively. Therefore, considering the sources of perturbation errors [15], $P_i \approx I_n - (Q_i R_i)((Q_i R_i)^T (Q_i R_i))^{-1} (Q_i R_i)^T$. Because $Q_i$ columns are orthogonal unit vectors, simplifying $R_i^T = R_i^T Q_i^T Q_i$ thanks to the $Q_i^T = Q_i^{-1}$ property with the rule that the transpose of a product is the product of the transpose in the reverse order, $(Q_i R_i)^T = R_i^T Q_i^T$ eases the notation of $P_i$ to $P_i \approx I_n - Q_i R_i (R_i^T R_i)^{-1} R_i^T Q_i^T$. Equivalently, the inverse of a product is the product of the inverses in the reverse order, so assuming that $R_i R_i^{-1} = (R_i^T)^{-1} R_i^T = I_n$, the simplified formula of the projection matrix onto the nullspace of $A_i$ is $P_i \approx I_n - Q_i Q_i^T$.

Considering the $m \gg n$ case, further decompositions may produce an excessive quantity of matrices, which is disadvantageous in the distributed environment due to the substantial task overhead time compared to its computational work time. Therefore, setting $A_j \subset A$, $j = 1, 2, 3, ..., J \in \mathbb{N}$ non-square full rank matrices $A_j \in \mathbb{R}^{l \times n}$, where $m > l > n$ and $I > J$ with doing full QR factorization on $A_j$, results in getting $(l \times l)$ unitary matrix multiplier where its last $(l \times (l - n))$ partition is multiplied by 0. It is far more efficient to get the first $n$ orthogonal columns using a reduced QR factorization form [16] instead.

$$A_j = Q_j \begin{bmatrix} R_j \\ 0 \end{bmatrix} = \begin{bmatrix} Q_{1_j} & Q_{2_j} \end{bmatrix} \begin{bmatrix} R_j \\ 0 \end{bmatrix} = Q_{1_j} R_j, \qquad j = 1, 2, ..., J \tag{1}$$

$\hat{x}_j(0) = R_j^{-1}((Q_{1_j})^T b_j)$ is the initial solution estimate vector of $A_j x_j = b_j$. Inverting the $R_j$ matrix is still a costly operation. Although, using the fact that the upper triangular $R_j$ matrix has one possible non-zero coefficient factor $r_{j_{n,n}}$ on the $n$-th row and column, the $n$-th component of $\hat{x}_j(0)$ can be quickly found using $b_j$ with the $n$-th vector of the $(Q_{1_j})^T$ matrix denoted as $q_{j_n}$.

$$\hat{x}_j(0)_n = \frac{q_{j_n} b_j}{r_{j_{n,n}}}, \qquad j = 1, 2, ..., J \tag{2}$$

Similarly, in each $R_j^{-1}$ matrix, entries $r_j^*$ are directly referenced with $r_j$ vector spaces in a way that the multiplication factors of the upper rows can be backwardly substituted [17] using previously calculated results $r^*_{j_{c-1,c}} \approx \frac{-r_{j_{c-1,c}}}{r_{j_{c-1,c-1}} r_{j_{c,c}}}$, where $c = n, n-1, ..., 1$ is a column number of $R_j$. Thus, there exists the ability for each $p$-th component of $\hat{x}_j(0)$ vector to be calculated recursively.

$$\hat{x}_j(0)_p = \frac{q_{j_p} b_j - \sum_{k=p+1}^{n} r_{j_{p,k}} x_k}{r_{j_{p,p}}}, \qquad (3)$$
$$p = n-1, n-2, ..., 1, \qquad j = 1, 2, ..., J$$

The computational complexity of inverting the $R_j$ matrix using the Gauss–Jordan elimination algorithm [18] is $\mathcal{O}(n^3)$. Even if optimized CW-like algorithms [19] are used alternatively, none of them would be as fast as backward substitution, which takes $\mathcal{O}(n^2)$.

$Q_{1_j}$ is the semi-orthogonal matrix [20], where $Q_{1_j} Q_{1_j}^T$ is the orthogonal projection onto its column space, and the projection matrix onto the nullspace of $A_j$ is remapped, implying isometry of Euclidean space.

$$P_j \mapsto I_n - Q_{1_j}^T Q_{1_j}, \qquad j = 1, 2, ..., J \qquad (4)$$

The average of the initial results across the computing nodes is simply the arithmetic average of the $J$ estimate results.

$$\bar{x}(0) = \frac{\sum_{k=1}^{J} \hat{x}_k(0)}{J} \qquad (5)$$

Having defined the $\gamma$ and $\eta \in (0, 1)$ parameters, updating the estimates and averaging the solutions through the $T$ number of epochs is done as defined in classical Accelerated Projection-Based Consensus [7].

$$\hat{x}_j(t+1) = \hat{x}_j(t) + \gamma P_j(\bar{x}(t) - \hat{x}_j(t)),$$
$$j = 1, 2, ..., J, \qquad t = 0, 1, ..., T-1 \qquad (6)$$

$$\bar{x}(t+1) = \frac{\eta}{J} \sum_{k=1}^{J} \hat{x}_k(t+1) + (1-\eta)\bar{x}(t),$$
$$t = 0, 1, ..., T-1 \qquad (7)$$

It is essential to identify what can be parallelized at a given stage of execution to minimize waiting processing time. Complete denotation is introduced in Algorithm 1.

**Algorithm 1:** Distributed Accelerated Projection-Based Consensus Decomposition

**Input:** Single or multiple full rank sparse coefficient matrices concatenated to $A$ with its corresponding concatenated constant terms vectors $b$, parameters $\eta$, $\gamma$, a number of partitions as $J$ and a number of epochs as $T$.

**Output:** Averaged solution vector $\bar{x}$ to a global system of linear or non-linear equations.

1. **Initialization:** Decompress $J$ submatrices from $A$ and $J$ subvectors from $b$ on worker nodes, propagate $I_n$ matrix.
2. Do QR decomposition (1) on $J$ matrices in parallel.
3. Concurrently find the initial solution (2) (3) for $J$ matrices with projection matrix (4) onto their nullspace.
4. Average the solution (5) over the nodes.
5. **for** $t = 0$ to $T-1$ **do**
6. In parallel, get the $t$-th solution update (6) for $J$ matrices.
7. Update the average solution (7) over the nodes.
8. **end for**

# 3. The Algorithm Implementation

Dask [9] is a flexible library for parallel computing in Python [10]. The processing workflow is scheduled in the form of a computational graph, exploited by the Dask scheduler, and lazily evaluated by Dask workers. An instance of such a graph is shown in Figure 1. Operations on computing nodes defined in Algorithm 1 are done using SciPy [21] and NumPy [22] built-in array object representations with their assortment of functional routines. The proposed implementation has been carried out using the virtualized computing environment of the Centre of Informatics Tricity Academic Supercomputer and networK.

```
import dataclasses

import dask.array
import dask.delayed
import dask.distributed
import numpy as np
import scipy as sp


@dataclasses.dataclass
class DistributedAPCDecomposition:
    scheduler_node_ip_address: str
    worker_nodes_ip_addresses: list[str]
    coefficient_matrix_path: str
    constant_terms_vector_path: str
    number_of_partitions: int
    epochs: int
    eta: float
    gamma: float

    def setup_dask_cluster_connection(self) -> dask.distributed.client.Client:
        hosts = [self.scheduler_node_ip_address] + self.worker_nodes_ip_addresses
        cluster = dask.distributed.SSHCluster(hosts)
        client = dask.distributed.Client(cluster)
        return client
```

```python
@staticmethod
@dask.delayed(nout=2, traverse=False)
def create_submatrices(
        A: sp.sparse._csr.csr_matrix, b: np.ndarray,
        partition_number: int, chunk_size: int) -> tuple[np.ndarray, np.ndarray]:

    if (partition_number+2)*chunk_size > len(b):
        return A[partition_number*chunk_size:, :].toarray(), b[partition_number*chunk_size:]
    return A[partition_number*chunk_size: (partition_number+1)*chunk_size, :].toarray(), \
           b[partition_number*chunk_size: (partition_number+1)*chunk_size]

@staticmethod
@dask.delayed(nout=2, pure=True)
def qr_decomposition(submatrix: np.ndarray) -> tuple[np.ndarray, np.ndarray]:
    return sp.linalg.qr(submatrix, mode='economic')

@staticmethod
@dask.delayed(pure=True)
def initial_solution(Q: np.ndarray, R: np.ndarray, subvector: np.ndarray) -> np.ndarray:
    return sp.linalg.solve_triangular(R, Q.T @ subvector)

@staticmethod
@dask.delayed
def create_identity_matrix(A: np.ndarray) -> np.ndarray:
    return np.identity(A.shape[1])

@staticmethod
@dask.delayed(pure=True)
def projection(I: np.ndarray, Q: np.ndarray) -> np.ndarray:
    return I - Q.T @ Q

@staticmethod
@dask.delayed(pure=True)
def average_initial_solutions(x_initials: list[np.ndarray]) -> np.ndarray:
    return np.mean(x_initials, axis=0)

@dask.delayed(pure=True)
def update_solution(self, x: np.ndarray, x_average: np.ndarray, P: np.ndarray) -> np.ndarray:
    return x + self.gamma * P @ (x_average - x)

@dask.delayed(pure=True)
def average_solutions(self, x: list[np.ndarray], x_average: np.ndarray) -> np.ndarray:
    return self.eta * np.mean(x, axis=0) + (1 - self.eta) * x_average

def graph(self) -> np.ndarray:
    A = sp.io.mmread(self.coefficient_matrix_path).tocsr()
    b = sp.io.mmread(self.constant_terms_vector_path)
    I = self.create_identity_matrix(A)
    chunk_size = len(b) // self.number_of_partitions
    x = []
    P = []

    for partition_number in range(self.number_of_partitions):
        submatrix, subvector = self.create_submatrices(A, b, partition_number, chunk_size)
        Q, R = self.qr_decomposition(submatrix)
        P.append(self.projection(I, Q))
        x.append(self.initial_solution(Q, R, subvector))

    x_average = self.average_initial_solutions(x)

    for _ in range(self.epochs):
        x[:] = [self.update_solution(x[index], x_average, P[index])
                for index in range(self.number_of_partitions)]
        x_average = self.average_solutions(x, x_average)

    return x_average

def run(self) -> np.ndarray:
    client = self.setup_dask_cluster_connection()
    graph = dask.delayed(self.graph())
    result = graph.compute()
    client.shutdown()
    return result
```

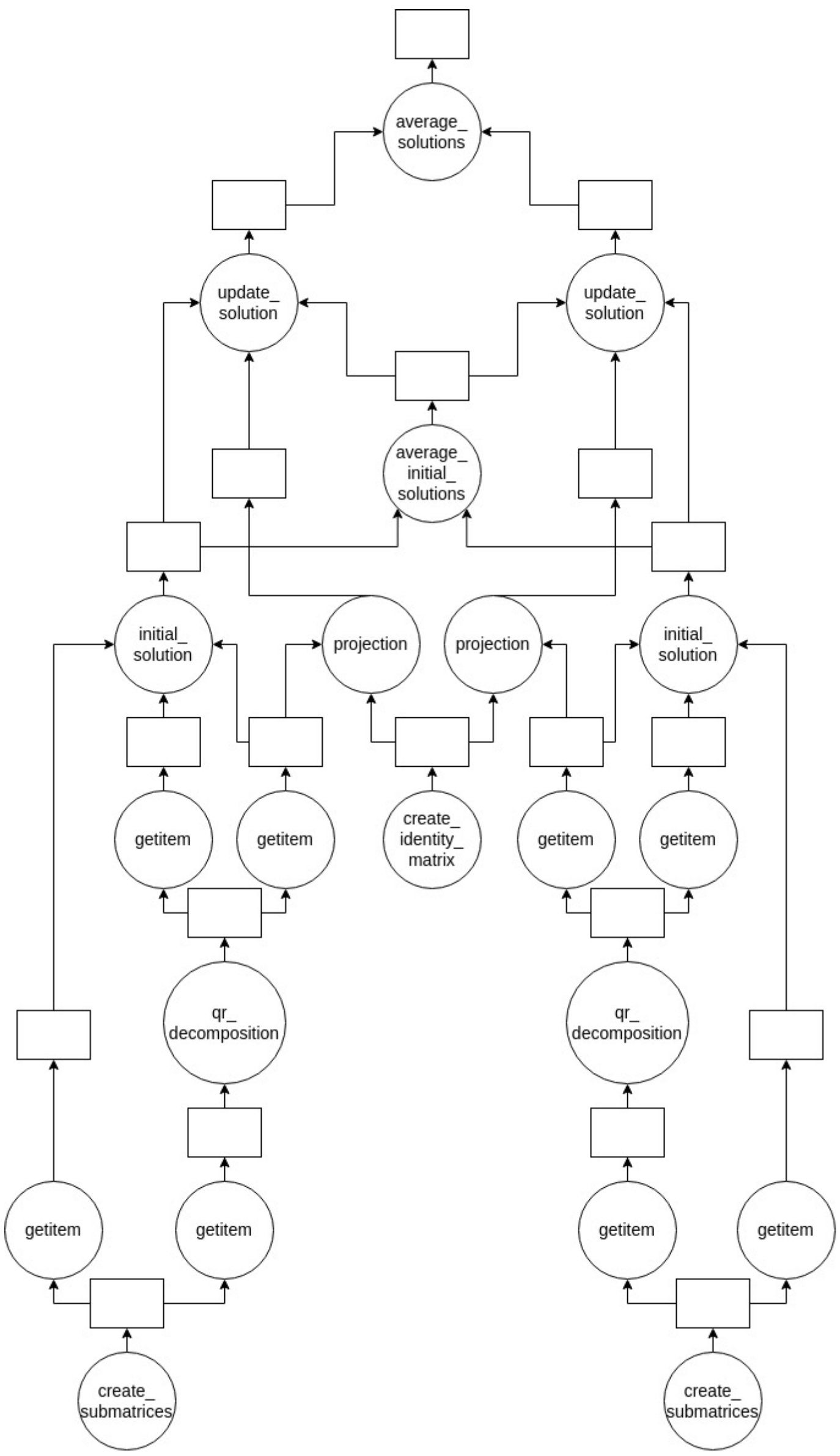

Figure 1. Computational graph representation performing a single-iteration computation of a two-partitioned input dataset

# 4. Convergence Behavior

Taking the full rank coefficient matrix $A \in \mathbb{R}^{n \times n}$, constant terms vector $b \in \mathbb{R}^n$, and pre-calculated $x \in \mathbb{R}^n$ as the most accurate solution vector to a system of equations, we can create an augmented $D_A \in \mathbb{R}^{m \times n}$ matrix with the correlative $D_b \in \mathbb{R}^m$ vector, linearly combined from $A$ and $b$, respectively.

$$\begin{bmatrix} A \\ D_A \end{bmatrix} \hat{x} = \begin{bmatrix} b \\ D_b \end{bmatrix} \tag{8}$$

Let $J$ partitions out of $\begin{bmatrix} A \\ D_A \end{bmatrix}$ be divisible such that each partition initialized on workers is not rank deficient, and $\frac{m+n}{J} \geq n$. Knowing that the original system of equations is consistent, the Algorithm's 1 output is defined as a multivariable function $f$ meeting the assumption $\lim_{T \to \infty} f(\begin{bmatrix} A \\ D_A \end{bmatrix}, \begin{bmatrix} b \\ D_b \end{bmatrix}, \eta, \gamma, J, T) = x$. Hence, the mean squared error (MSE) [23] between the $n$-th vector components of $\hat{x}$ and $x$ should be lower. Tests have been conducted on various sizes of Schenk_IBMNA datasets from the SuiteSparse Matrix Collection [24] with heuristically chosen parameters.

Due to approximations, the decomposed APC mean squared error of the initial solution should always be greater than in classical APC, where the initial solution is assumed to be found using matrix inverses. Both solutions converge to approximately the same level of minima in Figure 2 and are compared with the baseline Distributed Gradient Descent (DGD) [5] error, keeping in mind more firm possible divergencies in the decomposed APC after some $t$-th epoch. Statistics of multiple runs on different datasets are arranged in Table 1.

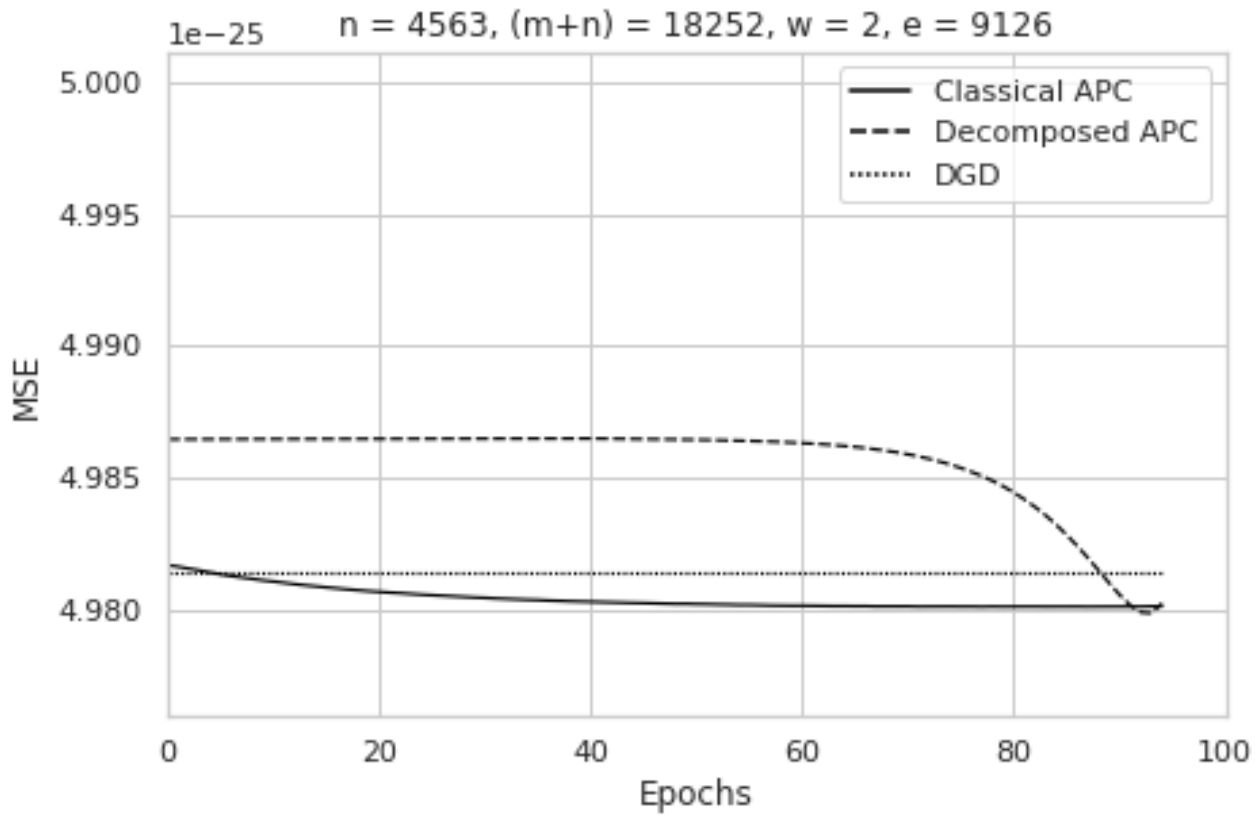

Figure 2: Average squared difference between the estimated $\hat{x}$ value and $x$ value over the number of epochs. The test was performed on the modified c-27 dataset [24], where $n$ = # of variables, $(m + n)$ = # of equations, $w$ = # of workers, $e$ = # of equations per worker

Table 1. Total execution times measurement where each algorithm approximately reaches its minima, run on $w = 2$, 4-core, single-threaded workers using independent input datasets

| $A$ matrix shape | $T$ number of epochs | Averaged processing wall time | | Acceleration |
|---|---|---|---|---|
| | | Classical APC | Decomposed APC | |
| $(9308 \times 2327)$ | 80 | $\approx 12.2s$ | $\approx 9.87s$ | $\approx 1.24$ |
| $(15188 \times 3797)$ | 70 | $\approx 31.6s$ | $\approx 21.2s$ | $\approx 1.49$ |
| $(18252 \times 4563)$ | 95 | $\approx 52.3s$ | $\approx 34.4s$ | $\approx 1.52$ |
| $(21284 \times 5321)$ | 85 | $\approx 74s$ | $\approx 44s$ | $\approx 1.68$ |
| $(37084 \times 9271)$ | 175 | $\approx 379s$ | $\approx 212s$ | $\approx 1.79$ |

# 5. Example

Consider the $(18252 \times 4563)$ coefficient matrix $A$ having $\mu = 0.013$, $\sigma = 24.31$ and the sparsity level of 99.85, together with $(18252 \times 1)$ constant terms vector $b$ taken from Schenk_IBMNA datasets [24].

$$A = \begin{bmatrix} 0.0718430341 & 0 & \dots & 0 & 0 \\ 0 & 0.0330022355 & \dots & 0 & 0 \\ \dots & \dots & \dots & \dots & \dots \\ 0 & 0 & \dots & -4e-08 & 0 \\ 0 & 0 & \dots & 0 & -4e-08 \end{bmatrix},$$

$$b = \begin{bmatrix} -0.01265569 \\ 0.0027185 \\ -0.00044829 \\ \dots \\ 0 \\ 0 \\ 0 \end{bmatrix} \tag{9}$$

Referring to Figure 1, the processing output of the graph is $(4563 \times 1)$ solution vector having $\mu \approx -0.0027$ and $\sigma \approx 0.0763$.

$$x = \begin{bmatrix} -0.10751059 \\ -0.15494688 \\ -0.03007913 \\ \dots \\ 0.00175945 \\ 0.0004609 \\ 0.00060833 \end{bmatrix} \quad (10)$$

The magnitude of the mean absolute error (MAE) [25] between the initial solution and the solution after the one iteration should be relatively small($< 1e-08$)for such $A$ and $b$. However, it suggests some direction of change. The output of subsequent iterations can be adjusted by configuring the $\eta$ and $\gamma$ hyperparameters.

# 6. Conclusion

The suggested variant of Accelerated Projection-Based Consensus [7] relying on factorization approximations is notably faster in terms of processing than its classical model, assuming matrix inversions. Both variants converge. The proposed variant is insignificantly less accurate and burdened with less stability. However, due to its ease and speed of implementation, it is a reasonable alternative for distributed optimization problems. It can be used as a baseline model in numerous real-world applications, especially in engineering.

## Acknowledgments

Greatest appreciation to the Accelerated Projection-Based Consensus [7] contributors for their support of the open access movement as well as the Centre of Informatics Tricity Academic Supercomputer and networK for providing computing power on the Tryton supercomputer.

## References

[1] Alexander Burton. OLS (Linear) Regression, pp. 509–514. Wiley, 08 2021.

[2] W. Keith Nicholson. Linear Algebra with Applications. Lyryx Learning Inc., Calgary, Alberta, Canada, 2020. Book version 2021A.

[3] Arthur Stanley Goldberger et al. Econometric theory. New York: John Wiley & Sons., 1964.

[4] Julien Simon. Large language models: A new Moore's law?, Oct 2021.

[5] Brian Swenson, Ryan Murray, Soummya Kar, and H Vincent Poor. Distributed stochastic gradient descent: Nonconvexity, nonsmoothness, and convergence to local minima. arXiv preprint arXiv:2003.02818, 2020.

[6] Ermin Wei and Asuman Ozdaglar. Distributed alternating direction method of multipliers. In 2012 IEEE 51st IEEE Conference on Decision and Control (CDC), pp. 5445–5450, 2012.

[7] Navid Azizan-Ruhi, Farshad Lahouti, Salman Avestimehr, and Babak Hassibi. Distributed solution of large-scale linear systems via accelerated projection-based consensus, 2017.

[8] Eman Shaikh, Iman Mohiuddin, Yasmeen Alufaisan, and Irum Nahvi. Apache spark: A big data processing engine. pp. 1–6, 11 2019.

[9] Matthew Rocklin. Dask: Parallel computation with blocked algorithms and task scheduling. pp. 126–132. Python in Science Conference, 01 2015.

[10] Guido van Rossum. Python programming language. In USENIX Annual Technical Conference, 2007.

[11] V. Klema and A. Laub. The singular value decomposition: Its computation and some applications. IEEE Transactions on Automatic Control, 25(2):164–176, 1980.

[12] Stephen Andrilli and David Hecker. Chapter 1 – vectors and matrices. In Stephen Andrilli and David Hecker, editors, Elementary Linear Algebra (Fifth Edition), pp. 1–83. Academic Press, Boston, fifth edition, 2016.

[13] E.K.P. Chong and S.H. Zak. An Introduction to Optimization. Wiley-Interscience Series in Discrete Mathematics and Optimi. Wiley, 2004.

[14] Walter Gander. Algorithms for the QR-decomposition. Seminar für Angewandte Mathematik: Research report, 1980.

[15] Erik Vleck. On the error in the product QR decomposition. SIAM J. Matrix Analysis Applications, 31:1775–1791, 01 2010.

[16] L.N. Trefethen and D. Bau. Numerical Linear Algebra. Other Titles in Applied Mathematics. Society for Industrial and Applied Mathematics (SIAM, 3600 Market Street, Floor 6, Philadelphia, PA 19104), 1997.

[17] Alberto Moreira. Cs557a: Solving linear systems of equations. 2000.

[18] Steven C. Althoen and Renate McLaughlin. Gauss-Jordan reduction: A brief history. The American Mathematical Monthly, 94(2):130–142, 1987.

[19] Zhikuan Zhao, Jack K Fitzsimons, Michael A Osborne, Stephen J Roberts, and Joseph F Fitzsimons. Quantum algorithms for training Gaussian processes. Physical Review A, 100(1):012304, 2019.

[20] Daniel Povey, Gaofeng Cheng, Yiming Wang, Ke Li, Hainan Xu, Mahsa Yarmohammadi, and Sanjeev Khudanpur. Semi-orthogonal low-rank matrix factorization for deep neural networks. In Interspeech, pp. 3743–3747, 2018.

[21] Pauli Virtanen, Ralf Gommers, Travis E. Oliphant, Matt Haberland, Tyler Reddy, David Cournapeau, Evgeni Burovski, Pearu Peterson, Warren Weckesser, Jonathan Bright, Stéfan J. van der Walt, Matthew Brett, Joshua Wilson, K. Jarrod Millman, Nikolay Mayorov, Andrew R.J. Nelson, Eric Jones, Robert Kern, Eric Larson, C J Carey, İlhan Polat, Yu Feng, Eric W. Moore, Jake VanderPlas, Denis Laxalde, Josef Perktold, Robert Cimrman, Ian Henriksen, E.A. Quintero, Charles R. Harris, Anne M. Archibald, Antônio H. Ribeiro, Fabian Pedregosa, Paul van Mulbregt, and SciPy 1.0 Contributors. SciPy 1.0: Fundamental Algorithms for Scientific Computing in Python. Nature Methods, 17:261–272, 2020.

[22] Charles R. Harris, K. Jarrod Millman, Stéfan J. van der Walt, Ralf Gommers, Pauli Virtanen, David Cournapeau, Eric Wieser, Julian Taylor, Sebastian Berg, Nathaniel J. Smith, Robert Kern, Matti Picus, Stephan Hoyer, Marten H. van Kerkwijk, Matthew Brett, Allan Haldane, Jaime Fernández del Río, Mark Wiebe, Pearu Peterson, Pierre Gérard-Marchant, Kevin Sheppard, Tyler Reddy, Warren Weckesser, Hameer Abbasi, Christoph Gohlke, and Travis E. Oliphant. Array programming with NumPy. Nature, 585(7825):357–362, September 2020.

[23] Alexei Botchkarev. A new typology design of performance metrics to measure errors in machine learning regression algorithms. Interdisciplinary Journal of Information, Knowledge, and Management, 14:045–076, 2019.

[24] Timothy A. Davis and Yifan Hu. The university of Florida sparse matrix collection. ACM Trans. Math. Softw., 38(1), Dec 2011.

[25] Gary Brassington. Mean absolute error and root mean square error: which is the better metric for assessing model performance? In EGU General Assembly Conference Abstracts, EGU General Assembly Conference Abstracts, pp. 3574, April 2017.